\documentclass{article}

\usepackage{arxiv}

\usepackage[utf8]{inputenc} 
\usepackage[T1]{fontenc}    
\usepackage{hyperref}       
\usepackage{url}            
\usepackage{booktabs}       
\usepackage{amsmath}
\usepackage{amsfonts}       
\usepackage{nicefrac}       
\usepackage{microtype}      
\usepackage{lipsum}		
\usepackage{graphicx}
\usepackage{natbib}
\usepackage{doi}

\usepackage{algorithm} 
\usepackage[noend]{algpseudocode}

\usepackage{adjustbox}
\usepackage{graphicx}
\usepackage{multicol}
\usepackage{multirow}
\usepackage{harmony}
\usepackage{scalerel}
\newcommand{\lcirc}[1]{\scaleobj{.6}{\mbox{\Kr{#1}}}}

\usepackage{xpatch}
\xapptocmd\normalsize{%
 \abovedisplayskip=12pt plus 3pt minus 9pt
 \abovedisplayshortskip=0pt plus 3pt
 \belowdisplayskip=12pt plus 3pt minus 9pt
 \belowdisplayshortskip=7pt plus 3pt minus 4pt
}{}{}

\title{A Variant RSA Acceleration with Parallelization}

\author{
    Jun-Jie LIU\\
	Division of Science and Technology\\
	BNU‐HKBU United International College \\
	Zhuhai 519085, China\\
	\texttt{junjieliu@uic.edu.cn} \\
	\And
    Kang-Too TSANG\thanks{Corresponding author}  \\
	Guangzhou Gaozhuan Information Technology Limited \\
	Guangzhou,  Guangdong, China \\
	\texttt{k2zeng@icloud.com} \\
	\And
	Yu-Hui Deng\\
	Division of Science and Technology\\
	BNU‐HKBU United International College \\
	Zhuhai 519085, China\\
	\texttt{ivandeng@uic.edu.cn} \\
}

\renewcommand{\shorttitle}{A Variant RSA Acceleation with Parallelization}

\hypersetup{
pdftitle={A Variant RSA Acceleration with Parallelization},
pdfsubject={cs.DC},
pdfauthor={Junjie-Jie LIU, Kang-Too TSANG, Yu-Hui DENG},
pdfkeywords={Parallel Computing, GPGPU, CUDA Implementation, Asymmetric Cryptosystem, RSA Algorithm},
}

\begin{document}
\maketitle

\begin{abstract}
The standard RSA relies on multiple big-number modular exponentiation operations and longer key-length is required for better protection. This imposes a hefty time penalty for encryption and decryption. In this study, we analyzed and developed an improved parallel algorithm (PMKRSA) based on the idea of splitting the plaintext into multiple chunks and encrypt the chunks using multiple key-pairs. The algorithm in our new scheme is so natural for parallelized implementation that we also investigated its parallelization in a GPU environment. In the following, the structure of our new scheme is outlined, and its correctness is proved mathematically. Then, with the algorithm implemented and optimized on both CPU and CPU+GPU platforms, we showed that our algorithm shortens the computational time considerably, and it has a security advantage over the standard RSA as it is invulnerable to the common attacks. Finally, we also proved the feasibility of using our algorithm to encrypt large files through simulation. The results show that over the set of file size: 1 MB, 10 MB, 25 MB, 50 MB, 100 MB, the average encryption and decryption time of the CPU version is 0.2476s and 9.4476s, and for the CPU+GPU version, it is 0.0009s and 0.0618s, respectively.
\end{abstract}


\section{Introduction}
With the rapid development of information network transmission technology, the requirement for the independence, security, and confidentiality of data transmission plays a vital role in the matter of E-commerce, telecommunication, and cloud computing. Cryptography is the main research field that studies methods for protecting information and data. In modern cryptosystem(non-quantum), the asymmetric cryptosystem is essential for today's internet as its ability in providing a secure key transformation method in non-secure communication channels, such as  Transport Layer Security (TLS), Secure Sockets Layer (SSL), Pretty Good Privacy (PGP). The main ideas of the public key system are:

\begin{enumerate}
    \item[1.] The public key is used for the encryption, and it can be disseminated
    \item[2.] The private key is used for the decryption, and it only kept by the owner
\end{enumerate}

The fundamental of the public-key cryptographic system (PKC) is an one-way function's generation: for two set $M$ and $C$, the function $f(\cdot)$ where $f: M \mapsto C$ is a one-way function if each $x$ belongs to $X$ and it is easy to calculate $f(X)$ and for most of $y$ belongs to Y, it is difficult to calculate $y = f(x) \mapsto x$. For the PKC, the decryption process is finding the inverse function, $f(x) \mapsto x$, where it should be difficult without the private key. 

In our study, we focus on one of the most well-celebrated asymmetric algorithms: RSA \cite{rivest1978method} which is proposed by Ron Rivest, Adi Shamir, and Leonard Adleman in 1978. We present an novel approach in a heterogeneous-structure machine by using CPU and GPU in separated tasks to accelerate both encryption and decryption with multiple key-pairs form different message chunks. This novel approach can also guarantee information security under some certain known attacks.

The structure of this paper is as follows. In section \ref{Introduction: Basic RSA}, we would introduce the standard RSA algorithm and section \ref{Introduction: Problems} has listed the problems we are facing when using the RSA. In section \ref{sec:Literiture Review}, we would introduce some variant RSA algorithms and also the mainstream parallel architectures using in scientific computing. In section \ref{sec:Novel Scheme}, a  parallelized variant RSA scheme has been proposed, and we also present the security analysis for the novel scheme in section \ref{Novel Scheme:Security-Analysis}. Finally, in section \ref{sec:Result}, we show the simulation results and then do the comparison with standard RSA.

\subsection{Standard RSA Cryptosystem}\label{Introduction: Basic RSA}    

The standard RSA algorithm includes three procedures:
\begin{enumerate}
    \item[1.] \textbf{Key Initialization:} 
    \begin{enumerate}
        \item[a.] Randomly choosing two large prime numbers $p \ \& \  q$, calculate $N = p \times q$ and $\phi(N) = (p-1)  \times (q-1)$, where $N$'s length is the key length.
        \item[b.] Choose $e \in \mathbb{Z}^*_{\phi(N)}$, where $e$ is the public exponent.
        \item[c.] Calculate $d = e^{-1}$ in $\mathbb{Z}^*_{\phi(N)} \Rightarrow ed \equiv 1 \pmod{\phi(N)}$, where $d$ is the private exponent
    \end{enumerate}
    \item[2.] \textbf{Message Encryption:} \par Assumes the plaintext is $M \in \mathbb{Z}_{N-1}$, The ciphertext is $C$: $C = M^e \pmod{N}$
    \item[3.] \textbf{Message Decryption:} \par By calculating $M' = C^d \pmod{N}$, we can get the plaintext.
\end{enumerate}

The proof of correctness of RSA is intuitive:
$M' = C^d \pmod{N} = M^{ed} \pmod{N} = M^{1 + k\phi(N)} \pmod{N} = M \times M^{k \phi{N}} = M \times (M^{\phi(N)})^k = M$; If $M$ is a relatively prime to $N$, according to Euler's Theorem, $M^{\phi(N)} \equiv 1 \pmod{N} \Rightarrow M' = M$ \label{correctness-RSA}

\subsection{Current trend in the Standard RSA Scheme}\label{Introduction: Problems}

The security of the RSA algorithm is based on the mathematical problem that the large number decomposition is difficult\cite{zhang2011research}. In 2000 and 2010, the RSA-512(base 2) and RSA-768(base 2) has been successfully factorized. People realize the small key-length is no longer secure and try to use the longer key to protect their information: the National Institute of Standards and Technology(NIST) suggests using at least $2048$-bits RSA keys in 2018 \cite{barker2018recommendation}. This comes up another question: with the length of key growth, the time of key generation, encryption, and decryption will also increase, which may cause a time-consuming problem for the user. In recent years, multiple types of research have been done in reducing the operation time, and these research can be summarized in these three aspects:

\begin{enumerate}
    \item [1.] Reducing the time complexity of modular exponentiation    
    \item [2.] Using Parallel computing to accelerate the computation
    \item [3.] Using the specific hardware to accelerate the encryption and decryption
\end{enumerate}

\noindent The second problem of using RSA is how to resist the crypt-analysis. For decades, cryptanalyst and researchers have conducted analysis and research on RSA for its vulnerability. Here are some typical attacks which is widely used in crypt-analysis to evaluate the security of a scheme:

\begin{enumerate}
    \item[1.] \textbf{Ciphertext Only Attack (COA)} or Known Ciphertext Attack (KCA) is a direct attack by using brute-force. Intuitively, the attacker is trying to break the system by factoring the large integers.
    \item[2.] \textbf{Known Plaintext Attack (KPA)} \cite{cryptoeprint:2012:588} is that the attacker has the ability to collect at least one random sample of plaintext and the corresponding ciphertext. In this case, the attacker can build up a set of plaintext and ciphertext: $S = \{(P_1, C_1), (P_2, C_2), \dots\}$ and break the private exponent and Modulus based on the set.
    \item[3.] \textbf{Chosen Plaintext Attack (CPA)}'s idea is similar to the KPA, however, instead of randomly getting information, under the Chosen Plaintext Attack, the attacker has the ability to specify which plaintext and corresponding will be known.
    \item[4.] \textbf{Chosen Ciphertext Attack (CCA)} assumes Alice sends the ciphertext $C$ to Bob with $C = P^e \pmod{N}$. With CCA assumption, the attacker knows the public exponent $e$ and he can get the decrypted message (except $C$'s) of chosen ciphertext. Therefore, $\forall X \in \mathbb{Z}^*_N$, the attacker can get $C_x = X^e \pmod{N}$ and get the decrypted message$M_a$. What's more, $M_a = (C \times C_x)^d \Rightarrow (C \times X^e)^d \Rightarrow C^d X = PX \pmod{N}$. Since $X$ and $n$ are co-prime, it is easy to calculate the modular multiplication inverse. 
\end{enumerate}

\section{Related Works}\label{sec:Literiture Review}
\subsection{Variant RSA Algorithms}\label{Literiture Review: Variant}

There are two kinds of approaches: multi-prime RSA proposed by Collins \cite{collins1998public} and multi-power RSA proposed by  Takagi \cite{takagi1998fast}. Both of these variants implemented the Chinese Remainder Theorem\cite{katz2007mathematics}: Given pairwise co-prime positive integers $n_1, n_2, \dots, n_k$ and arbitrary integers $d_1, d_2, \dots, d_k$, the system of linear congruence: 
\begin{equation}
    \left\{
    \begin{array}{lcr}
    &x \equiv d_1 \pmod{n_1} \\
    &x \equiv d_2 \pmod{n_2} \\
    &\vdots \\
    &x \equiv d_k \pmod{n_k} \\
    \end{array}
    \right.
\end{equation}

The system has a integer solution, and if $X$ and $Y$ satisfies the equation, we can draw a conclusion that $X \equiv Y \pmod{N}$, $N = \prod_{i=1}^k{n_i}$. The CRT-RSA substitutes the private exponent $d$ by factored private-key exponent pair: $d_p \in \mathbb{Z}_{p} \ \& \ d_q \in \mathbb{Z}_{q}$, and the the decryption procedure has shown in equation \ref{eq:CRT-RSA}:
\begin{equation}
    \left\{
    \begin{array}{lcr}
    d_p &\equiv d \pmod{p-1} \\
    d_q &\equiv d \pmod{q-1} \\
    \end{array}
    \right.
\Rightarrow 
    \left\{
    \begin{array}{lcr}
    m_p &= C^d \pmod{p} \\
    m_q &= C^d \pmod{q} \\
    h   &= q^{-1} (m_1 - m_2) \pmod{p} \\ 
    m'  &= m_2 + h \times q
    \end{array}
    \right.
\label{eq:CRT-RSA}
\end{equation}

For the multi-prime RSA, the main difference is in key generation, by using $b$ distinct primes $[p_1, p_2, \dots, p_b]$ and each of these primes are $\lfloor \frac{n}{b} \rfloor$, where $n$ is the base-2 length of RSA modulus. The multi-prime RSA's speedup over standard RSA \cite{boneh2002fast} is: 

\begin{equation}
    \frac{2 \cdot (n/2)^3}{b \cdot (n/b)^3} = \frac{b^2}{4}
\end{equation}

The Multi-power RSA is an alternative fast variant RSA also referenced as Takagi's RSA \cite{takagi1998fast}. The algorithm's goal is to accelerate the process of encryption and decryption by using multiple small prime numbers. $N = p^{b-1} q$, where $p$ and $q$ are two distinct prime numbers with the length equal to $\lfloor \frac{n}{b} \rfloor$ and $b \geq 3$. The exponents' generation can intuitively follow the standard RSA procedure or by using the $ed \equiv 1 \pmod{N}$.

The Dual RSA \cite{sun2007dual} is variant RSA algorithm based on the idea of sharing the same exponents and generating two pairs of prime numbers. The public key is: $(e, N_1, N_2)$ and the private key is $(d, p_1, p_2, q_1, q_2)$. By using the CRT, it provides 4 sub-private exponents:

\begin{equation}
    \left\{
    \begin{array}{lr}
    d &\equiv e^{-1} \pmod{N_1} \\
    d &\equiv e^{-1} \pmod{N_2} \\
    \end{array}
    \right.
\Rightarrow 
    \left\{
    \begin{array}{lr}
    d_{p_1} &\equiv d \pmod{p_1-1} \\
    d_{q_1} &\equiv d \pmod{q_1-1} \\
    d_{p_2} &\equiv d \pmod{p_2-1} \\
    d_{q_2} &\equiv d \pmod{q_2-1} \\
    \end{array}
    \right.
\label{eq:Dual-RSA-CRT}
\end{equation}

With the above Dual RSA-CRT equations, the algorithm can reduce both the computation time and the storage of keys, which is suitable for blind signature and authentication.

\subsection{Acceleration of RSA based on Parallelization}\label{Literiture Review: Parallel}

General-purpose computing with GPU (GPGPU) has became practical and popular after Tompson et al. 's research \cite{thompson2002using}, where they developed a general framework in solving high time-complexity problems including matrix multiplication and 3-SAT problem with the graphic card. Cook et al. \cite{cook2005cryptographics} shows that GPU can be used in cryptographic computing.  Since most of the Cryptography-related algorithms are based on arithmetic operations, these algorithms are typically compute-intensive. Hence, the parallelization of the sequential algorithm is an approach on the computational optimization. 

Both symmetric and asymmetric cryptography algorithm has been implemented with GPU. For symmetric cryptography, the Implementation of the Advanced Encryption Standard(AES) and Data Encryption Standard(DES) have been done by various researchers with different schemes and platforms \cite{manavski2007cuda, cook2005cryptographics, harrison2007aes, che2008performance}. In 2012, Hong Zhang\cite{zhang2012comparison} and his team have done the analysis of parallel RSA and compared the results between CUDA-based\footnote{CUDA: Parallel computing platform and programming model developed by NVIDIA for GPGPU} and Multi-core CPU-based algorithm. The experiments show that GPU implementation of modular exponentiation can achieve more than 45x acceleration. In 2010, Li and Liu \cite{li2010design} has proposed a improved RSA algorithm EAMRSA, Encrypt Assistant Multi-Prime RSA aiming to decrease the time of modular exponentiation by reducing modules and private exponent. The results of their algorithm shows that the EA1M4RSA(2048-bits to 3072-bits) is around 7 times faster than the Standard RSA with OpenMP \footnote{An API supports multi-platform shared-memory parallel programming in C/C++ and Fortran} Implementation. N.Cruz-Cortés et al.\cite{cruz2016gpu} presents the implementation of Residue Number System(RNS) with GPU can drastically reduce the computation time, and MA Ayub\cite{ayub2019parallelized} shows the experiments on RSA with a high-performance computing cluster (HPCC) to evaluate the acceleration.

\section{Parallelized Multi-key RSA (PMKRSA) Scheme}\label{sec:Novel Scheme}
\subsection{The architecture of the PMKRSA Scheme}\label{Novel Scheme: Architecture}

The standard RSA cannot encrypt the plaintext longer than the modulo $N$'s. In practice, we can split the plaintext into multiple chunks $M = [m_1, m_2,\dots, m_k]$. for $\forall i \in \mathbb{Z}_k$, $m_i$ can be encrypted via the RSA as its length is no larger than $N$'s. What's more, splitting the message can decrease the computing time as well. In previous scheme\cite{gupta2019matrix}, after the plaintext splitting, they use the same key-pair to encrypt and decrypt the plaintext. In our proposed parallelized multi-key RSA scheme, we reshape the plaintext vector into a $i \times j$ matrix (let $i \times j$ not less than $k$) and used different key-pairs for different rows:

$$M = \begin{bmatrix} 
    m_{11} & m_{12} & \dots & m_{1j}\\
    m_{21} & \ddots &  & \vdots \\
    \vdots & &  \ddots & \\
    m_{i1} & \dots &  & m_{ij}
\end{bmatrix}$$

The multiple key-pairs' components are illustrated as: 
\begin{enumerate}
    \item[1.] multiple public exponents: $\mathcal{E} = \left[e_1, e_2, \dots, e_{i}\right]^T$
    \item[2.] multiple private exponents: $\mathcal{D} = \left[d_1, d_2, \dots, d_{i}\right]^T$
    \item[3.] modulo: $\mathcal{N} = \left[N_1, N_2, \dots, N_{i}\right]^T$ 
\end{enumerate}

The difference between PMKRSA and RSA is in the step of encryption: we use a random number matrix $\mathcal{R}$ in the encipherment. For each plaintext chunk, we generate a corresponding random element and do the blinding\cite{yan2007cryptanalytic}. The random matrix can increase the ciphertext's randomness.

$$\mathcal{R} = \begin{bmatrix} 
    r_{11} & r_{12} & \dots & r_{1j}\\
    r_{21} & \ddots &  & \vdots \\
    \vdots & &  \ddots & \\
    r_{i1} & \dots &  & r_{ij}
\end{bmatrix}$$

For $\forall a \in \mathbb{Z}_i$ and $\forall b \in \mathbb{Z}_j$, each ciphertext chunk $m_{ab}$ corresponds to one random element $r_{ab} \in \mathbb{Z}^{+}$. The sender will generate the random matrix for each time of encryption and make sure it is cryptographically secure. The encryption process becomes\footnote{We define $\lcirc{e}$ and $\lcirc{d}$ as the exponentiation in encryption and the exponentiation in decryption respectively for the matrix form plaintext, $\circ$ is the symbol of Hadamard product.}:

\begin{align}
C^{*} &= \mathcal{R} \circ M^{\mathcal{E}} \pmod{\mathcal{N}} \\ &= 
\begin{bmatrix} r_{11} & r_{12} & \dots & r_{1j}\\
                    r_{21} & \ddots &  & \vdots \\
                    \vdots & &  \ddots & \\
                    r_{i1} & \dots &  & r_{ij}
\end{bmatrix} \circ
\begin{bmatrix} m_{11} & m_{12} & \dots & m_{1j}\\
                    m_{21} & \ddots &  & \vdots \\
                    \vdots & &  \ddots & \\
                    m_{i1} & \dots &  & m_{ij}
\end{bmatrix} \lcirc{e} 
\begin{bmatrix}
            e_1 \\
            e_2 \\
            \vdots \\
            e_{i}
\end{bmatrix} \pmod{\mathcal{N}}\\
&= 
\begin{bmatrix} (r_{11}\times m_{11})^{e_1} \pmod{N_1} & \dots & (r_{1j} \times m_{1j})^{e_1} \pmod{N_1}\\
                    (r_{21} \times m_{21})^{e_2} \pmod{N_2} & \ddots &  \vdots \\
                    \vdots &   & \vdots \\
                   (r_{i1} \times m_{i1})^{e_{i}} \pmod{N_i} & \dots & (r_{ij} \times m_{ij})^{e_{i}} \pmod{N_{i}}
\end{bmatrix} = 
\begin{bmatrix} c_{11} & c_{12} & \dots & c_{1j}\\
                    c_{21} & \ddots &  & \vdots \\
                    \vdots & &  \ddots & \\
                    c_{i1} & \dots &  & c_{ij}
\end{bmatrix} \\
C_R &= R^{\mathcal{E}} \pmod{\mathcal{N}} \\ &= 
\begin{bmatrix} r_{11} & r_{12} & \dots & r_{1j}\\
                    r_{21} & \ddots &  & \vdots \\
                    \vdots & &  \ddots & \\
                    r_{i1} & \dots &  & r_{ij}
\end{bmatrix} \lcirc{e} 
\begin{bmatrix}
            e_1 \\
            e_2 \\
            \vdots \\
            e_{i}
\end{bmatrix} \pmod{\mathcal{N}}\\
&= 
\begin{bmatrix} r_{11}^{e_1} \pmod{N_1} & \dots & r_{1j}^{e_1} \pmod{N_1}\\
                    r_{21}^{e_2} \pmod{N_2} & \ddots &  \vdots \\
                    \vdots &   & \vdots \\
                   r_{i1}^{e_{i}} \pmod{N_i} & \dots & r_{ij}^{e_{i}} \pmod{N_{i}}
\end{bmatrix} = 
\begin{bmatrix} c_{r_{11}} & c_{r_{12}} & \dots & c_{r_{1j}}\\
                    c_{r_{21}} & \ddots &  & \vdots \\
                    \vdots & &  \ddots & \\
                    c_{r_{i1}} & \dots &  & c_{r_{ij}}
\end{bmatrix}
\end{align}

The encrypted message includes $C^{*}$ and $C_R$. For the decryption:

\begin{align}
R^{\prime} &= C_R^{\mathcal{D}} \pmod{\mathcal{N}} \\ &= 
\begin{bmatrix} c_{r_{11}} & c_{r_{12}} & \dots & c_{r_{1j}}\\
                c_{r_{21}} & \ddots &  & \vdots \\
                \vdots &   &  \ddots & \\
                c_{r_{i1}} & \dots &  & c_{r_{ij}}
\end{bmatrix} \lcirc{d}
\begin{bmatrix}
            d_1 \\
            d_2 \\
            \vdots \\
            d_{i}
\end{bmatrix} \pmod{\mathcal{N}}\\
&= 
\begin{bmatrix} c_{r_{11}}^{d_{1}} \pmod{N_1} & \dots & c_{r_{1j}}^{d_1} \pmod{N_1}\\
                    c_{r_{21}}^{d_{1}} \pmod{N_2} & \ddots &  \vdots \\
                    \vdots &   & \vdots \\
                   c_{r_{i1}}^{d_{i}} \pmod{N_i} & \dots &  c_{r_{ij}}^{d_{i}} \pmod{N_{i}}
\end{bmatrix} = 
 \begin{bmatrix} r^{\prime}_{11} & m'_{12} & \dots & r^{\prime}_{1j}\\
                    r^{\prime}_{21} & \ddots &  & \vdots \\
                    \vdots & &  \ddots & \\
                    r^{\prime}_{i1} & \dots &  & r^{\prime}_{ij}
\end{bmatrix} \\
M^{\prime} &= (C^*)^{\mathcal{D}} \times {R^{\prime}}^{-1} \pmod{\mathcal{N}} \\ &= 
\begin{bmatrix} r^{\prime}_{11} & m'_{12} & \dots & r^{\prime}_{1j}\\
                    r^{\prime}_{21} & \ddots &  & \vdots \\
                    \vdots & &  \ddots & \\
                    r^{\prime}_{i1} & \dots &  & r^{\prime}_{ij}
\end{bmatrix} \circ
\begin{bmatrix} 
                    c_{11} & c_{12} & \dots & c_{1j}\\
                    c_{21} & \ddots &  & \vdots \\
                    \vdots & &  \ddots & \\
                    c_{i1} & \dots &  & c_{ij}
\end{bmatrix} \lcirc{d}
\begin{bmatrix}
            d_1 \\
            d_2 \\
            \vdots \\
            d_{i}
\end{bmatrix} \pmod{\mathcal{N}}\\
&= 
\begin{bmatrix} c_{11}^{d_{1}} \times {r^{\prime}_{11}}^{-1} \pmod{N_1} & \dots & c_{1j}^{d_1} \times {r^{\prime}_{1j}}^{-1}\pmod{N_1}\\
                c_{21}^{d_{2}} \times {r^{\prime}_{21}}^{-1}\pmod{N_2} & \ddots &  \vdots \\
                \vdots &   & \vdots \\
                c_{i1}^{d_{i}} \times {r^{\prime}_{i1}}^{-1}\pmod{N_i} & \dots &  c_{ij}^{d_{i}} \times {r^{\prime}_{ij}}^{-1} \pmod{N_{i}}
\end{bmatrix}
=
 \begin{bmatrix} m^{\prime}_{11} & m^{\prime}_{12} & \dots & m^{\prime}_{1j}\\
                    m^{\prime}_{21} & \ddots &  & \vdots \\
                    \vdots & &  \ddots & \\
                    m^{\prime}_{i1} & \dots &  & m^{\prime}_{ij}
\end{bmatrix}
\end{align}

The proof of correctness of PMKRSA is shown below, For each message chunks $m_{ab} \in \mathbb{Z}_{N_a}$:

\begin{align*}
    m^{'*}_{ab} &= c_{ab}^{d_{a}} \pmod{N_a} \\
                 &= ((m_{ab} \times r_{ab})^{e_{a}} \pmod{N_a})^{d_{a}} \pmod{N_a} \\
                 &= m_{ab} \times r_{ab} \\
    m^{'}_{ab}  &= m^{'*}_{ab} \times r^{-1}_{ab} \\
                 &= m^{'}_{ab} \times r_{ab} \times r^{-1}_{ab} \\
                 &= m^{'}_{ab}
\end{align*}

The Algorithm \ref{PMKRSA} is the pseudo-code of PMKRSA For each communication, the sender first selects the fixed-length random matrix $\mathcal{R}$, and then encrypts the plaintext and the $\mathcal{R}$ as $(C^{*}, C_R)$. When the receiver receive the message, the first step is to decrypt the $C^r$ and use the decrypted $\mathcal{R}'$ to decrypt the $C^{*}$.

\begin{algorithm}[H]     
    \caption{Parallelized Multiple-Key RSA Scheme (PMKRSA)} 	
    \label{PMKRSA} 	
    \begin{algorithmic}[1]
        \Procedure{Initialization}{The plaintext matrix is $i \times j$, length of key: $n$}
        \For {$a \in [1, i]$}
            \State{Choose two prime number $p$ and $q$, where $p \neq q$ with $\lfloor{\frac{n}{2}}\rfloor$}
            \State{$N_a = p \times q$}
            \State{Choose $e_a \in \mathbb{Z}^*_{\phi(N_a)}$}
            \For {$b \in [1,j]$}
                \State{Choose a random number $r_{ab} \in \mathbb{Z}_{N_a}$}
                \State{$d_{a} = e_a^{-1} \in \mathbb{Z}^*_{\phi(N_a)}$}
            \EndFor
        \EndFor 
        \State {\textbf{Return:} Public Key: $[\mathcal{E}, \mathcal{N}]$ and Private Key: $[\mathcal{D}, \mathcal{N}]$}
        \EndProcedure 
        \Procedure{Encryption}{plaintext:$M$, Ciphertext:$C$, Public Exponent:$\mathcal{E}$} 		
        \For {$a \in [1, i]$}
            \For {$b \in [1, j]$}
                \Comment{The ``For" loops can be parallelized}
                \State{$c_{ab} = m_{ab}^{e_{a}} \pmod{N_a}$}
                \State{$c^*_{ab} = c_{ab} \times r_{ab}^{e_{a}} \pmod{N_a}$}
                \State{$c_{r_{ab}} = r_{ab}^{e_a} \pmod{N_a}$}
            \EndFor
        \EndFor 
        \State {\textbf{Return} Ciphertext: ($C^*$,$C_{R}$)}
        \EndProcedure 		
        \Procedure{Decryption}{Ciphertext: ($C^*$,$C_{R}$), Private Exponent: $\mathcal{D}$}
        \For {$a \in [1, i]$}
            \For {$b \in [1, j]$}
                \Comment{The ``For" loops can be parallelized}
                \State{$r^{\prime}_{ab} = c_{r_{ab}}^{d_{a}} \pmod{N_a}$}
                \State{${m^{\prime}_{ab}}^* = {c^{*}_{ab}}^{d_{a}} \pmod{N_a}$}
                \State{$m^{\prime}_{ab} = {m^{\prime}_{ab}}^* \times {r^{\prime}_{ab}}^{-1} \pmod{N_a}$}
            \EndFor
        \EndFor		
        \State {\textbf{Return} Decryption message $M'$} 		
        \EndProcedure 	
    \end{algorithmic} 
\end{algorithm}

\subsection{Big Integer Arithmetic and Parallel Platform}\label{Novel Scheme: BigInteger}

Working with big/large integers is necessary for any RSA-related cryptosystem implementation. There are multiple libraries provide BIG-NUM structure, for example, OpenSSL is a famous crypto-library \footnote{for more detail in OpenSSL: https://github.com/openssl/openssl} which provides a robust and sustainable "big integer" implementation called:``BIGNUM"; The GNU Multiple Precision Arithmetic Library(GMP) \footnote{GMP: https://gmplib.org/} is a open-source arithmetic library that operates on integers and floating-point numbers for arbitrary precision. These two libraries are both support C/C++. We use GMP library in our experiment since there is no limitation of precision by using the GMP library, theoretically.

The key operation in the RSA algorithm is the modular exponentiation of large integers: $y = x^e \pmod{n}$. To calculate the modular exponentiation, we can transform the operation into a series of modular multiplication or modular squaring operations, named as multiply-and-square exponentiation. The following algorithm is multiply-and-square exponentiation (left-to-right method) \cite{moller2002improved} based on binary splitting. In each iteration, the square and multiplication are independent and the execution can be parallelized:

\begin{algorithm}[H]     
    \caption{Multiply-And Square Exponentiation (Base 2)} 	
    \label{alg:quik-exponentiation} 	
    \begin{algorithmic}[1]
        \Procedure{Multiply-And Square Exponentiation}{} 
        \State{Input: $\forall x, y, e, N \in \mathbb{Z}^{+}_0$}
        \State{Initiation: $y = 1, x = x \pmod{n}$}
        \For {$i \in [0, l)$, where $e = (e_l,e_{l-1},\dots, e_0)_2$ and $e_i \in \{0, 1\}$}
            \If {$e_i == 1$}
                \State {$y = (y \times x) \pmod{N}$}
            \EndIf
            \State {$x = x \times x \pmod{N}$}
        \EndFor
        \State{\textbf{Return} $y = y \times x \pmod{N}$}
        \EndProcedure 		
    \end{algorithmic} 
\end{algorithm}

In order to avoid the division operation in modular exponentiation, we implement the Montgomery algorithm, which is proposed by Peter Montgomery in 1985. The algorithm and detailed proof can be found in \cite{montgomery1985modular}, and the Montgomery multiplication is summarized in the Algorithm: \ref{alg:montgomery-multiplication}

\begin{algorithm}[H]     
    \caption{Montgomery Multiplication} 	
    \label{alg:montgomery-multiplication} 	
	\begin{algorithmic}[1]
	    \State {\textbf{Input:} $a, b, r, N$, where $r = 2^k$}
	    \State {\textbf{Output:} $\hat{x} \times \hat{y} \times r^{-1}\pmod{N}$}
        \State {$\hat{a} = a \times r \pmod{N}$}
        \State {$\hat{b} = b \times r \pmod{N}$}
        \State {$t = \hat{a} \times \hat{b}$}
        \State {$n' \equiv -N^{-1} \pmod{r}$}
        \State {$t = (t + m \times N) / r$}
        \If {$t \geq N$}
            \State{\textbf{return} t-N}
        \Else{:}
            \State{\textbf{return} t}
        \EndIf
	\end{algorithmic}
\end{algorithm}

To store the big integer in GPU, we allocate the same memory size in GPU and copy the corresponding memory information from CPU to GPU. By using this method, we can easily transfer the data into the GPU instead of designing a new structure and this can avoid the data conflict.

Parallel programming, also known as parallel computing, is an alternative to conventional serial computing. Rather than allowing a single instruction to be executed in a single operation time, it allows multiple instructions to be executed at the same time with multiple computational resources. There are three main models in parallel computing:

\begin{enumerate}
    \item[1.] Message-passing programming
    \item[2.] Shared-address programming
    \item[3.] Data-parallel programming
\end{enumerate}

\begin{figure}[H]
    \centering
    \includegraphics[width=12cm]{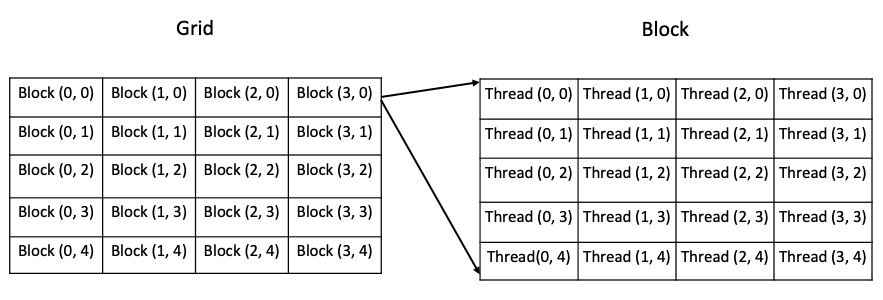}
    \caption{Grid, Block and Thread in CUDA}
    \label{fig:grid-block-thread}
\end{figure}

OpenMP is a commonly used shared memory programming API since it facilitates the development of parallel applications by using compiler directives directly. Furthermore, it can be easily deployed and used in different systems. CUDA is a parallel computing platform proposed by NVIDIA \cite{nvidia2006cuda} which allows users to operate a computation on the GPU with C/C++, Fortran, etc. As the graphic card contains thousands of core, a parallel program can be executed in multiple threads and these threads would be formed into a block. The threads in the same block can be synchronized or communicate through shared memory. Multiple blocks will then form a grid, an example is shown in Figure \ref{fig:grid-block-thread}. 

\subsection{Implementation in Heterogeneous Device}\label{Novel Scheme: Detail}
The GPU includes more computing cores, which are especially suitable for data-consistent computationally intensive tasks, such as the matrix multiplication. Although the CPU contains fewer computational cores, it more caches. This means it is suitable for control-intensive tasks. In addition, different to the GPU, CPU's threads are heavy-wight, and the context switching overhead is large. Therefore, the [CPU+GPU] - based heterogeneous computing platform can complement each other. The CPU is responsible for processing complex serial programs, while the GPU focuses on data-intensive parallel computing programs to maximize performance.

\begin{figure}[H]
     \centering
     \begin{minipage}[b]{0.49\textwidth}
         \centering
         \includegraphics[width=\textwidth]{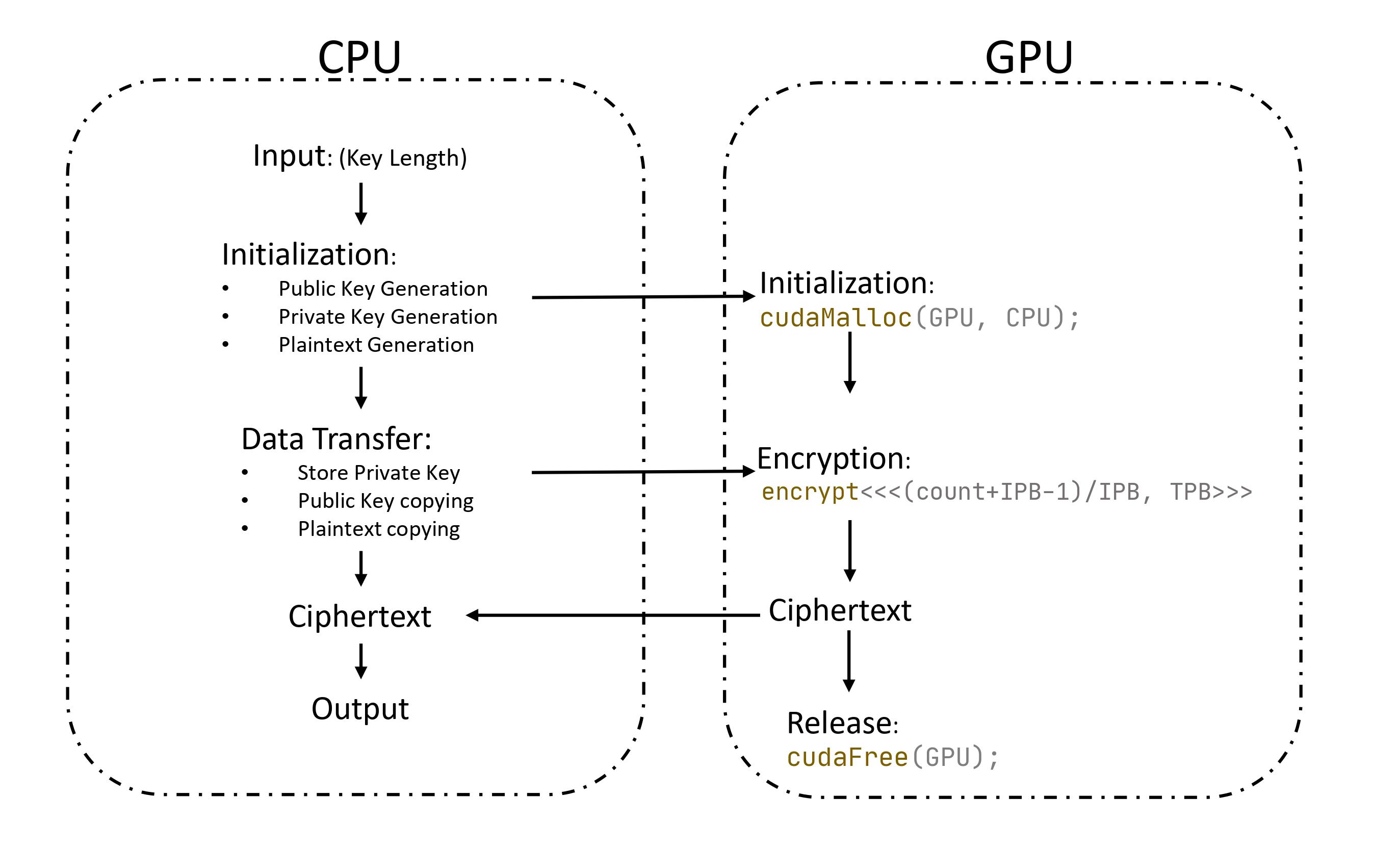}
         \caption{Encryption with Heterogeneous Architecture}
         \label{fig:CPU_GPU_Encrypt}
     \end{minipage}
     \hfill
     \begin{minipage}[b]{0.49\textwidth}
         \centering
         \includegraphics[width=\textwidth]{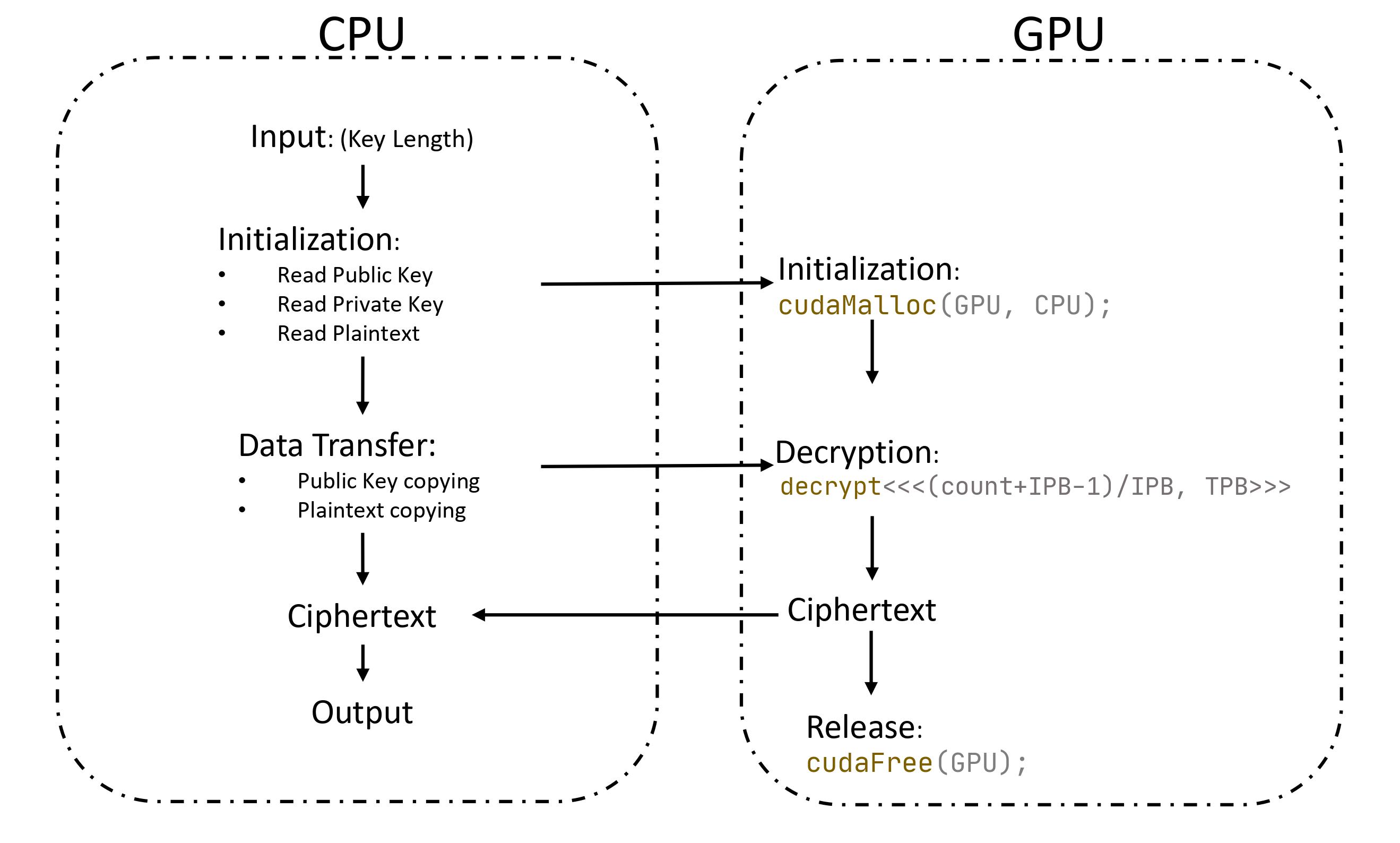}
         \caption{Decryption with Heterogeneous Architecture}
         \label{fig:CPU_GPU_Decrypt}
     \end{minipage}
\end{figure}
\begin{enumerate}
    \item[1.] For the initialization, First, the host would receive key length and plaintext. Next, the CPU would generate the key-pairs which is needed. As the scheme we proposed needs multiple key-pairs, which means if we generate keys sequentially, it would cost multiple time. Hence, we use OpenMP to parallelize the key generation operation. To ensure all the keys are unique, uncorrelated, and independent, instead of using \textit{/dev/urandom}, we choose \textit{/dev/random}\footnote{According to Linux Programmer's Manual, the \textit{/dev/random} only provide the output when it collects enough physical entropy} to generate the random seed, which is a LINUX cryptographically secure pseudorandom number generator with blocking mechanism\cite{gutterman2006analysis}. The \textit{/dev/random} may block if the entropy pool is exhausted, this can guarantee the keys are individual.
    \item[2.] Transfer and encryption/decryption: the private-key, public-key, plaintext and ciphertext are transferred from the host's memory to the GPU's memory while GPU performs the encryption/decryption operation. It is the key execution unit of a hybrid [CPU+GPU] system.
    \item[3.] Merge and output: the host needs to merge the encrypted/decrypted data. What is more, the hybrid [CPU+GPU] system is a pipeline architecture: each actuation unit is like each piece of equipment in an assembly line.The algorithm's parallelism is greatly improved by performing each sub-task intermittently.
\end{enumerate}

\subsection{Security Analysis of the Parallelized Multi-Key RSA}\label{Novel Scheme:Security-Analysis}

In this section, we discuss the security analysis of PMKRSA only considering the algorithmic attack without the physical and the hardware-related attack.

\subsubsection{Ciphertext-Only Attack (COA)}
The security of PMKRSA is based on two aspects: 
\begin{enumerate}
    \item[1.] The difficulty of the RSA factoring challenge, or the big integer factorization
    \item[2.] Security of private exponent (brute-force attack)
\end{enumerate}

For Big integer factorization, the length of $N$ we use is not less than 2048, which means the prime number we use is not less than 256-bits and the prime are not able to factorized by the Elliptic Curve method \cite{silverman1993practical, atkin1993finding,peralta1996faster}. Hence, the scheme can prevent from the integer factorization attack. 

For the security of private exponent, assume that all the public exponents are known by the attacker (for example, Our schemes has split the message into $k$ chunks with $i$ rows and $j$ columns, $\forall a\in \mathbb{Z}_i$ and $\forall b \in \mathbb{Z}_j$, $e_a = \mathbf{F}_4$ and the length of $N_a$ is $n$-bits) and we can prove that the attacker cannot search all the possible $r_{ab}$ and $d_{a}$: 
\begin{enumerate}
    \item[1.] To ensure the security of $r_{ab}$, the length of $r_{ab}$ cannot be too small. What's more, the multiplication of $r_{ab}$ and $m_{ab}$ cannot larger than $n$. Under this circumstance, we assume that the length of $len(r_{ab}) = len(m_{ab}) = \frac{n}{2}$.  Hence, The searching  of $\mathcal{R}$ is $2^{\frac{n \times k}{2}}$.
    \item[2.] The length of each private exponent $d_{a}$ is around $n$-bits and the searching space of $\mathcal{D}$ is $2^{n \times i}$.
\end{enumerate}

\subsubsection{Known plaintext Attack (KPA)}
We can also prove that PMKRSA is secure in against the KPA attack. The assumption of the attack is, the attacker know some of the plaintext and the corresponding public exponents $\{(m_{11}, c_{11}), (m_{12}, c_{12}), \dots\}$. With the PMKRSA, the attacker can only receive the ciphertext with blinding, that is $c^{*}_{ab} = m_{ab}^{e_{a}} \pmod{N_a} \times r_{ab}^{e_{a}} \pmod{N_{a}}$, and set up the mapping as: $\{(m_{11}, c^{*}_{11}), (m_2, c^{*}_{12}), \dots\}$. Therefore, the attacker cannot break the ciphertext and get the plaintext.

For the encrypted random matrix $C_R$, if the $C_R$ is deterministic, it cannot avoid the KPA attack as its encryption follows the standard RSA procedure. However, for each time of message transfer, the random matrix is re-generated by the sender. The attacker cannot deduce the $\mathcal{R}$ based on the previous information. Therefore, KPA on the random matrix $\mathcal{R}$ is infeasible.

\subsubsection{Chosen Plaintext Attack(CPA)}

The CPA attack is similar to the KPA, the different between them is the attacker with CPA can choose the ciphertext by itself. We can prove that the PMKRSA can against the CPA attack. It is because, the message has been encrypted in chunks and each ciphertext $c_{ab}$ is blinded with a random element $r_{ab}$. In this circumstance, the output $c_{ab}^* = c_{ab} \circ r_{ab}^{e_{a}} \pmod{N_{a}}$ is not deterministic anymore and plaintext-ciphertext mapping set is not unique. What's more, the random matrix $R$ is a disposable parameter and the $C_R$ cannot be break.

\subsubsection{Chosen Ciphertext Attack (CCA)}
CCA attack is computationally infeasible in PMKRSA. Following the above procedure (Algorithm \ref{PMKRSA}), with PMKRSA, because of the implementation of random number in encryption, the attacker can only get the $ \mathcal{R}_X \circ X \circ \mathcal{R}_C \circ C \pmod{\mathcal{N}}$ as the result. Although the random number $r_{ab}$ is not semantic secure, one $r_{ab}$ corresponds to only one ciphertext chunks, there are no leaking information of the next chunk. Therefore, the PMKRSA can against the Chosen Ciphertext Attack.

\section{Result and Discussion}\label{sec:Result}

In this session, we will describe the detail on our experiment environment, introduce the measurement unit we are using, and experiment results. Finally, we will have a discussion.

The Experiment device is a computing Server (Dell EMC R740), it contains a graphic card from NVIDIA. The tables below show the basic configuration of the device:
\begin{table}[H]
\centering
    \begin{tabular}{ll}
    \hline  
    CPU Processor: & Intel Xeon Silver 4114 (2.2GHz - 3.00GHz)\\
    \hline  
    System Memory: & 128GB DDR4 \\
    Hard Drive: & Intel Enterprise SSD 240GB \\
    Graphic Card: & NVIDIA Tesla P100 12GB (Pascal) \\
    Operating System: & Ubuntu 18.04 LTS \\
    CUDA Version: & CUDA 10.2 \\
    \hline
    \end{tabular}
\end{table}

Theoretically, We use algorithm time complexity to evaluate our algorithms. Assuming the length of the plaintext is $L$, the key-length is $n$ and the amount of chunks is $k$ and the bit size of a chunk is $\omega$, where $L, n, k, \omega \in \mathbb{Z}^+$. For the standard RSA algorithm with the textbook multiplication, the time complexity of encryption is $O(L \times n^2)$ and the decryption is $O(L \times n^3)$. By using the Montgomery reduction, it reduces the exponentiation time into $O(L \times \frac{3}{\omega^2} n^2)$ and $O(L \times \frac{3}{\omega^2} n^3)$ for the encryption and decryption respectively\cite{rawat2019novel}. The time complexity of PMKRSA is shown below:
\begin{enumerate}
    \item[a.] Encryption: $O((\frac{L}{k} + \frac{n}{2}) \times \frac{3}{\omega^2} n^2)$
    \item[b.] Decryption: $O((\frac{n}{2} + \frac{n}{2} \times \frac{L}{k}) \times \frac{3}{\omega^2} n^3)$
\end{enumerate}

In practice,the runtime of simulation is used to examine the algorithm's efficiency. We choose the ``speed-up rate'' as the evaluation standard:
$$\text{Speed Up} = \frac{\text{CPU Version Runtime}}{\text{[CPU+GPU] Hybrid Runtime}} = \frac{\text{Original scheme Runtime}}{\text{Novel Scheme Runtime}}$$

The first experiment is designed in two parts,  we implemented PMKRSA in CPU-only and [CPU+GPU] hybrid mode: for the CPU-only try to evaluate a parallel architecture for CPU only as well as for the hybrid version, we would like to demonstrate acceleration with such a heterogeneous architecture. The plaintext we use is 102400 bytes(B), which equals 100 kilobytes (KB). The performance of PMKRSA CPU version, PMKRSA [CPU+GPU] version  and standard RSA is shown Table \ref{result:CPU_scheme} and Table \ref{result:[CPU+GPU]_scheme}:

\begin{enumerate}
    \item[1.] Table \ref{result:CPU_scheme} shows the results of PMKRSA on OpenMP platform. For using the Configuration A, the average speedup of encryption and decryption is $27.75$ and $7.80$; for using the Configuration B, the average speedup of encryption and decryption is $36.11$ and $14.63$, respectively.
    \item[2.] Table \ref{result:[CPU+GPU]_scheme} shows the results of PMKRSA on with [CPU+GPU] architecture. First, Configuration D and standard RSA's speedup rate are $515.266$ (encryption) and $104.12$ (decryption). Comparing with the Configuration B and D, the speedup rate of hybrid version of encryption and decryption is $8.32$ and $3.86$, respectively.
\end{enumerate}

\begin{table}[H]
\centering
\resizebox{\textwidth}{!}{
\begin{tabular}{lllllll}
\hline
\multicolumn{7}{c}{PMKRSA using OpenMP (Intel Silver 4114)}                                   \\ \hline
          & \multicolumn{2}{c}{2 cores 4 threads with 16 keys (Configuration A)} & \multicolumn{2}{c}{4 cores 8 threads with 16 keys (Configuration B)} & \multicolumn{2}{c}{standard RSA}    \\ \hline
Length    & Encryption (sec)         & Decryption (sec)        & Encryption (sec)         & Decryption (sec)        & Encryption (sec) & Decryption (sec) \\
2048 bits & 0.0191                   & 0.4836                  & 0.0160                   & 0.2570                  & 0.2723           & 3.6436           \\
3072 bits & 0.0310                   & 1.5386                  & 0.0217                   & 0.8124                  & 0.5936           & 11.7555          \\
4096 bits & 0.0415                   & 3.5449                  & 0.0275                   & 1.8234                  & 1.0117           & 26.6927          \\
6144 bits & 0.0741                   & 10.7952                 & 0.0449                   & 5.7404                  & 2.1278           & 82.3200          \\
8192 bits & 0.1178                   & 24.7352                 & 0.0685                   & 12.8515                 & 3.5948           & 187.5091         \\
Average   & 0.0567                   & 8.2195                  & 0.0357                   & 4.2969                  & 1.5200           & 62.3842    \\ \hline
\end{tabular}}
\caption{Result of CPU Version with PMKRSA using OpenMP}
\label{result:CPU_scheme}
\end{table}

\begin{table}[H]
\centering
\resizebox{\textwidth}{!}{
\begin{tabular}{lllllll}
\hline
\multicolumn{7}{c}{PMKRSA using OpenMP \& CUDA (NVIDIA P100)}                              \\ \hline
          & \multicolumn{2}{c}{32 threads with 8 keys (Configuration C)} & \multicolumn{2}{c}{32 threads with 16 keys (Configuration D)} & \multicolumn{2}{c}{standard RSA}    \\ \hline
Length    & Encryption (sec)     & Decryption (sec)    & Encryption (sec)     & Decryption (sec)     & Encryption (sec) & Decryption (sec) \\
2048 bits & 0.0009               & 0.0611              & 0.0009               & 0.0611               & 0.2723           & 3.6436           \\
3072 bits & 0.0014               & 0.1434              & 0.0014               & 0.1419               & 0.5936           & 11.7555          \\
4096 bits & 0.0021               & 0.2818              & 0.0021               & 0.2763               & 1.0117           & 26.6927          \\
6144 bits & 0.0036               & 0.7911              & 0.0036               & 0.7873               & 2.1278           & 82.3200          \\
8192 bits & 0.0060               & 1.6714              & 0.0060               & 1.6665               & 3.5948           & 187.5091         \\
Average   & 0.0028               & 0.5897              & 0.0028               & 0.5866               & 1.5200           & 62.3842          \\ \hline
\end{tabular}}
\caption{Result of [CPU+GPU] Version with PMKRSA using OpenMP \& CUDA}
\label{result:[CPU+GPU]_scheme}
\end{table}

The second experiment is to compare the performance of PMKRSA in encrypting and decrypting the large file. The key length is 2048-bits and the tested files' size are: [1MB, 10MB, 25MB, 50MB, 100MB] with random generation. Table \ref{result:Big_file} includes the result of CPU version of PMKRSA with configuration B, [CPU+GPU] version with configuration D and widely used AES-256-CBC.

\begin{table}[h]
\resizebox{\textwidth}{!}{
\begin{tabular}{lllllll}
\hline
\multicolumn{7}{c}{Comparison Between PMKRSA and AES For Different File Sizes}                                                                                      \\ \hline
       & \multicolumn{2}{c}{CPU version with Configuration B} & \multicolumn{2}{c}{CPU+GPU version with Configuration D} & \multicolumn{2}{c}{OpenSSL AES-256-CBC} \\ \hline
Length & Encryption (sec)                  & Decryption (sec)                 & Encryption (sec)                & Decryption (sec)                & Encryption (sec)   & Decryption (sec)   \\
1MB    & 0.0139                            & 0.2567                           & 0.0009                          & 0.0617                          & 0.0310             & 0.0300             \\
10MB   & 0.0650                            & 2.5309                           & 0.0009                          & 0.0618                          & 0.1400             & 0.1390             \\
25MB   & 0.1527                            & 6.5128                           & 0.0009                          & 0.0612                          & 0.3200             & 0.3160             \\
50MB   & 0.3106                            & 12.8152                          & 0.0009                          & 0.0623                          & 0.6140             & 0.6160             \\
100MB  & 0.6958                            & 25.1223                          & 0.0009                          & 0.0622                          & 1.1970             & 1.2040  \\ \hline
\end{tabular}}
\caption{Comparison Between PMKRSA and AES For Different File Sizes}
\label{result:Big_file}
\end{table}

From These above three table, we can reveal some interesting relations between the CPU version's PMKRSA and [CPU+GPU] version's PMKRSA. First, we can conclude that the speed up ratio of encryption process between PMKRSA and standard RSA is much better than the decryption process. The reason of this phenomenon can be explained via the time complexity: the decryption process of PMKRSA has added an additional decryption process of $\mathcal{R}$ (including the modular exponentiation and the un-blinding process), which increases the complexity of the algorithm. The second point is, although the file size is increasing, the computation time of the [CPU+GPU] version is basically the same. This is because of the GPU's high throughput and the parallelization. With the high throughout, the GPU can receive the plaintext in a very short time, and with the parallelization, it can simultaneously compute multiple operations. 

\section{Conclusion}

In this work, we investigate multiple fast variant of RSA-based algorithms, for example, M-Prime RSA, M-Power RSA, and Dual RSA. We also provide a parallel computing idea applied to the RSA algorithm: Parallelized Multi-key RSA schemes which can be easily used in both multi-core CPU system and [CPU+GPU] heterogeneous architecture and can be widely used no matter in a individual PC or a lab-cluster. We prove that by using multiple-key scheme, the searching space of key-pairs is much larger than the standard RSA.We also proved that our schemes are secure in against common RSA-oriented attack. In general, The schemes have these advantages:
\begin{enumerate}
    \item[1.] They can overcome the problem of ``RSA can not encrypt arbitrarily long files" by splitting the message into multiple chunks
    \item[2.] The Modular exponentiation time usage decreased by implementing on the mainstream parallel platforms, which are user-friendly.
    \item[3.] Different message chunks use different key-pairs, which makes the entire algorithm more resistant to the brute-force attack. 
    \item[4.] PMKRSA introduces random numbers in the encryption process and we have proved that both PMKRSA can prevent from Known Plaintext Attack, Chosen Plaintext Attack and it is also the Chosen Ciphertext Attack. 
    \item[5.] The random matrix $\mathcal{R}$'s encryption and decryption may increase the algorithm's total complexity. However, with the parallelization and the hardware acceleration, we can overcome the time-consuming problem. The experiment result shows that by using the mainstream home computer configuration \footnote{We use only 2 cores and 4 cores to simulate the personal computer configuration. Usually, the frequency of server processor is lower than the PC's and the effect of PC's CPU acceleration will be more obvious.}, it can achieve a good acceleration effect.
\end{enumerate} 

These multi-key schemes can be used in many scenarios as it can encrypt the large file in a secure way. For example, the broadcast message transformation, where PMKRSA can avoid the Hastad's broadcast attack. It can also be used in full disk/file encryption without using other symmetric crypto-algorithms. 

\section{Acknowledgement}
This work is supported by Guangdong University Innovation and Enhancement Program Funds 2018KTSCX278; and also supported by internal research grant from Beijing Normal University-Hong Kong Baptist University-United International College (UIC) R201712, R201809 and R201810. The computation in this paper has been done on the USBC\footnote{USBC is short for: BNU-HKBU United International College Statistics Bayes Cluster} cluster provided by Beijing Normal University-Hong Kong Baptist University-United International College (UIC) Department of Statistics.

\bibliographystyle{unsrtnat}
\bibliography{references}

\appendix
\section{Glossary of Symbols}
\begin{table}[H]
    \centering
    \begin{tabular}{ll}
        \\ \hline
        Symbol                  &               Description \\ \hline
        $\mathbb{N}$ and $\mathbb{Z}^{+}$  &    Set of natural numbers: $\mathbb{N} = \mathbb{Z}^{+} = \{1, 2, 3, \dots\}$  \\          
        $\mathbb{Z}_0^{+}$      &               Set of non-negative numbers: $\mathbb{Z}_{0}^{+} = \{0, 1, 2, 3, \dots\}$  \\  
        $\mathbb{Z}$            &               Set of integers: $\mathbb{Z} = \{0, \pm n: n \in \mathbb{N}\}$ \\
        $\mathbb{Z}_N$          &               Residue classes modulo $N$: $\mathbb{Z}_N = \{0, 1, 2, \dots, N-1\}$ \\
        $\mathbb{Z}^*_N$        &               Multiplicative group: $\mathbb{Z}^*_N = \{a \in \mathbb{Z}_N: gcd(a, N)=1\}$ \\
        $\mathbf{F_i}$          &               the $i$th Fermat number \\
        $\lfloor x \rfloor$      &               the largest integer less than or equal to $x$ \\
        $p$ and $P$             &               Prime Number \\
        $q$ and $Q$             &               Prime Number \\
        $e$                     &               Public Exponent, used in RSA  \\
        $d$                     &               Private Exponent, used in RSA  \\
        $N$                     &               Modulo, used in RSA \\
        $\mathcal{E}$           &               Public Exponent, used in PMKRSA \\
        $\mathcal{D}$           &               Private Exponent, used in PMKRSA \\
        $\mathcal{R}$           &               Random Matrix, used in PMKRSA \\
        $\mathcal{N}$           &               Modulo, used in PMKRSA \\
        $M$                     &               plaintext, used in RSA and PMKRSA  \\
        $C$                     &               Ciphertext, used in RSA and PMKRSA  \\ 
        $k$                     &               The length of plaintext vectors\\\hline
    \end{tabular}
    \label{tab:symbols}
\end{table}

\end{document}